\documentclass[12pt, twocolumn]{aastex63}
\usepackage{physics}
\usepackage[normalem]{ulem}
\usepackage{amsmath, amssymb}
\usepackage{bbold}
\definecolor{beaver}{rgb}{0.62,0.51,0.44}

\newcommand{\LCDM}{\ensuremath{\Lambda\rm{CDM}}}

\begin{document}

\title{Model selection and parameter estimation using the iterative smoothing method}

\shortauthors{Koo et al.}
\email{hkoo@kasi.re.kr, shafieloo@kasi.re.kr, rkeeley@kasi.re.kr, blhuillier@yonsei.ac.kr}

\author[0000-0003-0268-4488]{Hanwool Koo}
\affil{Korea Astronomy and Space Science Institute, Daejeon 34055, Korea}
\affil{University of Science and Technology, 217 Gajeong-ro, Yuseong-gu, Daejeon 34113, Korea}

\author[0000-0001-6815-0337]{Arman Shafieloo}
\affil{Korea Astronomy and Space Science Institute, Daejeon 34055, Korea}
\affil{University of Science and Technology, 217 Gajeong-ro, Yuseong-gu, Daejeon 34113, Korea}

\author[0000-0002-0862-8789]{Ryan E. Keeley}
\affil{Korea Astronomy and Space Science Institute, Daejeon 34055, Korea}

\author[0000-0003-2934-6243]{Benjamin L'Huillier}
\affil{Yonsei University, Seoul 03722, Korea}

\date{\today}
\submitjournal{ApJ}

\begin{abstract}
We compute the distribution of likelihoods from the non-parametric iterative smoothing method over a set of mock Pantheon-like type Ia supernova datasets. We use this likelihood distribution to test whether typical dark energy models are consistent with the data and to perform parameter estimation. In this approach, the consistency of a model and the data is determined without the need for comparison with another alternative model. Simulating future WFIRST-like data, we study type II errors and show how confidently we can distinguish different dark energy models using this non-parametric approach.

\end{abstract}


\section{Introduction}\label{sec:intro}

For decades there has been no significant change in the concordance model of cosmology, $\LCDM$ ($\Lambda$ for the cosmological constant and CDM for the cold dark matter).
So far, it has been the most successful model, explaining various astronomical observations with remarkable simplicity. For instance, this model predicts the low-redshift dynamics of the Universe with only two parameters, the Hubble constant, $H_0$, and the matter density, $\Omega_{\rm m}$. 

Type Ia supernova (SN Ia) distance measurements have become one of the most important datasets of modern cosmology since they are standardizable candles and they directly measure the accelerating expansion of the Universe at late times \citep{riess-etal98,perlmutter-etal99}. Almost all previous SN Ia compilations including SuperNova Legacy Survey  \citep[SNLS,][]{sullivan-snlsc05}, Gold \citep{riess-etal07}, Union \citep{kowalski-etal08}, Constitution \citep{hicken-etal09}, Union2 \citep{amanullah-etal10}, Union2.1 \citep{suzuki-etal12}, Joint Light-curve Analysis \citep[JLA,][]{betoule-etal14} and Pantheon \citep{scolnic-etal18} have been shown to be consistent with the flat $\LCDM$ model\footnote{There are a few previous studies that find the SN Ia data can still allow for deviations from $\LCDM$ \citep{tutusaus-etal17,tutusaus-etal19,kim-etal18,TDE,kim-etal19}, or a weaker evidence for an accelerating Universe (e.g. see \cite{nielsen-etal16,colin-etal19} though these are disputed by \cite{rubin-hayden16,rubin-heitlauf20}, respectively.}. However, these consistency tests need to assume some parametrization or functional form, whatever their complexity   \citep[see][for a model-indepedent analysis of possible systematics in the Pantheon compilation]{lhuillier-etal19}.

Though $\LCDM$ may be consistent with low-redshift data, it is in conflict with the Swampland conjecture~\citep{SwampI,SwampII}, which states that, even as a low-energy effective theory, it cannot correspond to a full, high-energy theory of quantum gravity.
Presumably, this would indicate that eventually, future data should indicate the successes of $\LCDM$ will break down at some point. This may already be the case with the $H_0$ tension, a discrepancy between the present expansion rate measured directly from the Cepheid calibration of SN Ia distances~\citep{riess-etal19} and that rate inferred from the CMB~\citep{Planck18}.  To find a new model that will replace $\LCDM$, it can be useful to first use model independent methods to explore a wider set of expansion histories that are consistent with the data.  If certain features in these model independent expansion histories prove robust and significant, we can then build models around these features.

To this end, we use the non-parametric iterative smoothing method, introduced and improved by \cite{shafieloo-etal06,shafieloo07,shafieloo-clarkson10,shafieloo-etal18}, to reconstruct the distance modulus in a model-independent way. 
Further, we use the ``likelihood distribution'' to test the consistency between different dark energy models and the SN Ia data. This likelihood distribution can also be used to perform parameter estimation for each model.
We analyze the Pantheon compilation \citep{scolnic-etal18}, one of the most recent SN Ia compilations which provides distance moduli computed from standardized SALT2 \citep{guy-etal07,mosher-etal14} light-curve parameters.
Also, we simulate a mock Wide Field Infrared Survey Telescope\footnote{The name has changed to Nancy Grace Roman Space Telescope (Roman Space Telescope, RST) recently. However, we will use the previous name since it is still widely used in the field.} \citep[WFIRST,][]{green-etal12,spergel-etal15} SN Ia compilation and forecast those results. 

We introduce the methodology and calculate the rate of type I errors in Sec.~\ref{sec:pantheon} and calculate the type II error rate for future data in Sec.~\ref{sec:wfirst}. Finally, we present our discussions and conclusions in Sec.~\ref{sec:dis}.

\section{The Iterative smoothing method and likelihood distributions}\label{sec:pantheon}

In this section, we discuss a non-parametric iterative smoothing method used to reconstruct the distance modulus $\mu(z)$ from the data $\mu_i$ observed at redshifts $z_i$ and the expansion history of the Universe from an arbitrary initial guess, $\hat{\mu}_0(z)$. The distance modulus is reconstructed iteratively where the $n+1$ iteration, $\hat{\mu}_{n+1}(z)$, is calculated by
\begin{equation}\label{eqn:smooth}
\hat{\mu}_{n+1}(z) = \hat{\mu}_n(z) + \frac{\boldsymbol{\delta\mu_n}^T \cdot \mathbf{C^{-1}} \cdot \boldsymbol{W}(z)}{\mathbb{1}^T \cdot \mathbf{C^{-1}} \cdot \boldsymbol{W}(z)}
\end{equation}
where $\mathbb{1}^T=(1,\cdots,1)$, the weight $\boldsymbol{W}$ and residual $\boldsymbol{\delta\mu_n}$ denote
\begin{equation}\label{eqn:weight}
W_i(z)=\exp(-\frac{\ln^2\left(\frac{1+z}{1+z_i}\right)}{2\Delta^2})
\end{equation}
\begin{equation}\label{eqn:residual}
\boldsymbol{\delta\mu_n}|_i = \mu_i -  \hat{\mu}_n(z_i)
\end{equation}
and $\mathbf{C^{-1}}$ indicates the inverse of covariance matrix of the data. The smoothing width is set to $\Delta = 0.3$ following previous analyses in \cite{shafieloo-etal06, lhuillier-shafieloo17, lhuillier-etal18, koo-etal20}.

For the Pantheon dataset, the covariance matrix is the quadratic sum of the statistical light-curve fit uncertainty and the systematic uncertainties from the bias correction, calibration, Galactic extinction, light-curve model, and mass step correction. The systematic uncertainties also include systematic uncertainties caused by intrinsic scatter, peculiar velocity, redshift measurement, and stochastic gravitational lensing. \cite{scolnic-etal18} describes more details about the uncertainties.

We define the $\chi^2$ value of the reconstruction $\hat{\mu}_n(z)$ as
\begin{equation}\label{eqn:chi2}
\chi_n^2 = \boldsymbol{\delta\mu_n}^T \cdot \mathbf{C^{-1}} \cdot \boldsymbol{\delta\mu_n}.
\end{equation}
The iterative smoothing method has been used so far mainly to reconstruct a non-exhaustive sample of viable expansion history possibilities that can fit the data with a better likelihood than a specific threshold.
For instance, in \cite{lhuillier-etal18,shafieloo-etal18,koo-etal20} this method has been used to present a large sample of possibilities with viable smooth characteristics than can fit the data better than the best flat $\LCDM$ model. In this work, we attempt to tackle a different problem and seek to test the consistency of a particular model with the data by calculating a quantity we call the likelihood distribution, which is based on our reconstruction method and follows a frequentist statistical approach. 

The iterative smoothing method has some important characteristics that have been studied in previous works. 
For instance, at any iteration the reconstructed function fits the data better than the previous reconstruction. This is what the algorithm is designed to do.
Furthermore, after a large number of iterations the reconstructions converge to a unique solution independently of the choice of the initial guess model. 
In other words we can start the machinery with very different initial guesses that can have very different initial likelihoods to the data but after a large number of iterations the final reconstructions converges to the same solution with a unique likelihood. In this work we use the 1000th iteration of the iterative smoothing method, which is large enough to achieve this convergence (that generally occurs after a few hundreds of iterations)\citep{shafieloo07,lhuillier-etal18,shafieloo-etal18}. This allows us to understand what is the best likelihood we should expect to get from our algorithm, independent of the initial guess model.

\newpage
\subsection{Model Selection}
It is a generic feature of this iterative smoothing method that it will produce a function that has a better $\chi^2$ value than that of the best-fit model.  We want to be able to answer the question, how much better does this improvement have to be in order to be significant.  To do so, we follow the typical frequentist approach and make mock datasets where we know the true cosmology.  Applying the smoothing procedure to these mock datasets then allows us to see how often the smoothing procedure generates better fits by certain amounts (i.e. what is the distribution of the difference in $\chi^2$ between the smoothed function and that of the best-fit model).  This distribution is what we call the likelihood distribution.  
We ultimately want to derive a number $\Delta\chi^2_\text{95\%}$ (or similar) such that if the improvement between the iterative smoothing and the best-fit model is larger than $\Delta\chi^2_\text{95\%}$, then we conclude that the model is a bad fit to the data.

In other words, we want to find the improvement in $\chi^2$ achieved by the smoothing method, such that only 5\% of the time would the smoothing method achieve a better improvement than this value by random chance (a type I error rate of 5\%).  This $\Delta \chi^2$ value then corresponds to the 95\% confidence level (CL). 

To validate our methodology and calculate the 95\% CL, we generate 1000 mock Pantheon-like datasets.  
We make these 1000 mock datasets for each of three separate cosmology cases, $\LCDM$, Phenomenologically Emergent Dark Energy (PEDE)~\citep{li-shafieloo19,li-shafieloo20}, and Kink~\citep{corasaniti-copeland03}.
This is to check that the 95\% CL we calculate is largely independent of the cosmological model. 

\begin{figure*}
\centering
\includegraphics[width=0.45\textwidth]{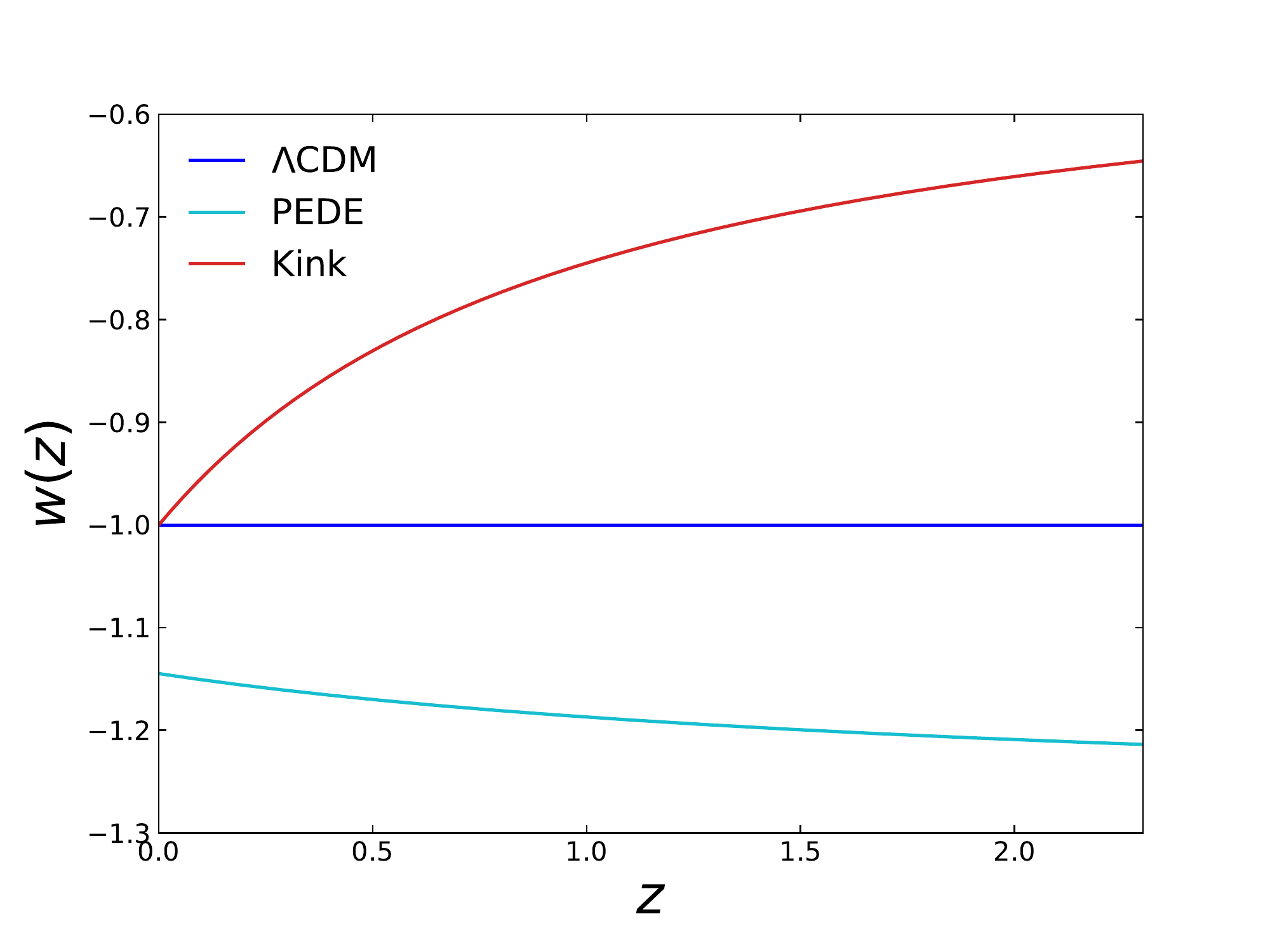}
\includegraphics[width=0.45\textwidth]{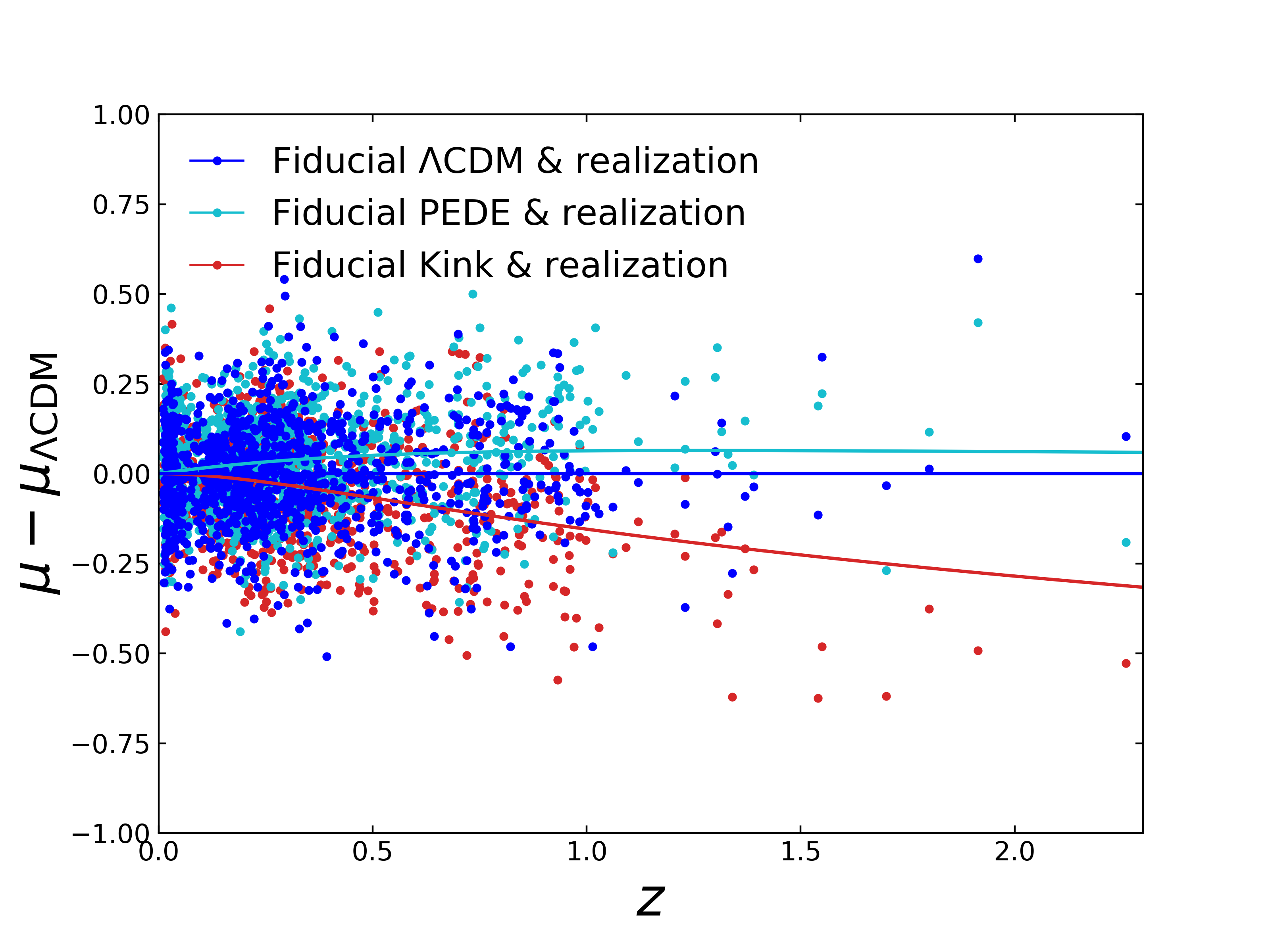}
\caption{\label{fig:w_dmu_models} (Left) Equation-of-state parameter of the $\LCDM$ (blue), PEDE (cyan), and Kink (red) models. (Right) Deviation of the distance moduli from that of $\LCDM$ model. We show a single random realization of the residuals distributed around the $\LCDM$, PEDE and Kink models with $H_0=70 \ \rm{km \ s^{-1} \ Mpc^{-1}}$ and $\Omega_{\rm m}=0.3$. The resampling is done using the Pantheon covariance matrix.}
\end{figure*}

In a flat FLRW universe with a dark energy component with equation-of-state $w(z)$, the luminosity distance can be written as
\begin{equation}\label{eqn:d_l_model}
d_L(z) = \frac{c}{H_0}(1+z)\int_{0}^{z}\frac{dz'}{E(z')}
\end{equation}
where the expansion history $E(z)$ is
\begin{equation}\label{eqn:h_z_model}
E^2(z) = \Omega_m(1+z)^3 + (1-\Omega_m)\exp(3\int_{0}^{z}\frac{1+w(z')}{1+z'}dz').
\end{equation}
In $\LCDM$, $w=-1$, but in general it can vary. For instance, The PEDE model, recently introduced by \cite{li-shafieloo19} and generalized by \cite{li-shafieloo20}, offers another example of the evolution of $w(z)$ where
\begin{align}
    w(z) & = -\tfrac{1}{3\ln 10} \left({1+\tanh\left[\log_{10}\,(1+z)\right]}\right)-1.
\end{align}
In the PEDE model, dark energy is absent in the past and acts as an emergent phenomena. We also consider the kink model where $w(z)$ is described by four parameters
\begin{equation}\label{eqn:w_kink}
w(z)\ =\ w_0\ +\ (w_{\infty}-w_0)\frac{1+\exp(\frac{a_c}{d_m})}{1+\exp(-\frac{a-a_c}{d_m})}\frac{1-\exp(-\frac{a-1}{d_m})}{1-\exp(\frac{1}{d_m})}.
\end{equation}
We choose $w_0=-1$, $w_{\infty}=-0.5$, $a_c=\frac{2}{3}$, and $d_m=1$ just as an example of an evolving dark energy model. The Kink model is an evolving dark energy model that allows a sharp transition in its equation of state. In this work, we use this specific set of parameters since they present such a transition and have been used earlier by \cite{holsclaw-etal10} and \cite{shafieloo-etal12}.
Fig.~\ref{fig:w_dmu_models} shows the equation of states for our three fiducial models and one of these mock Pantheon-like realizations from each of the three models.

\begin{figure}
\centering
\includegraphics[width=0.45\textwidth]{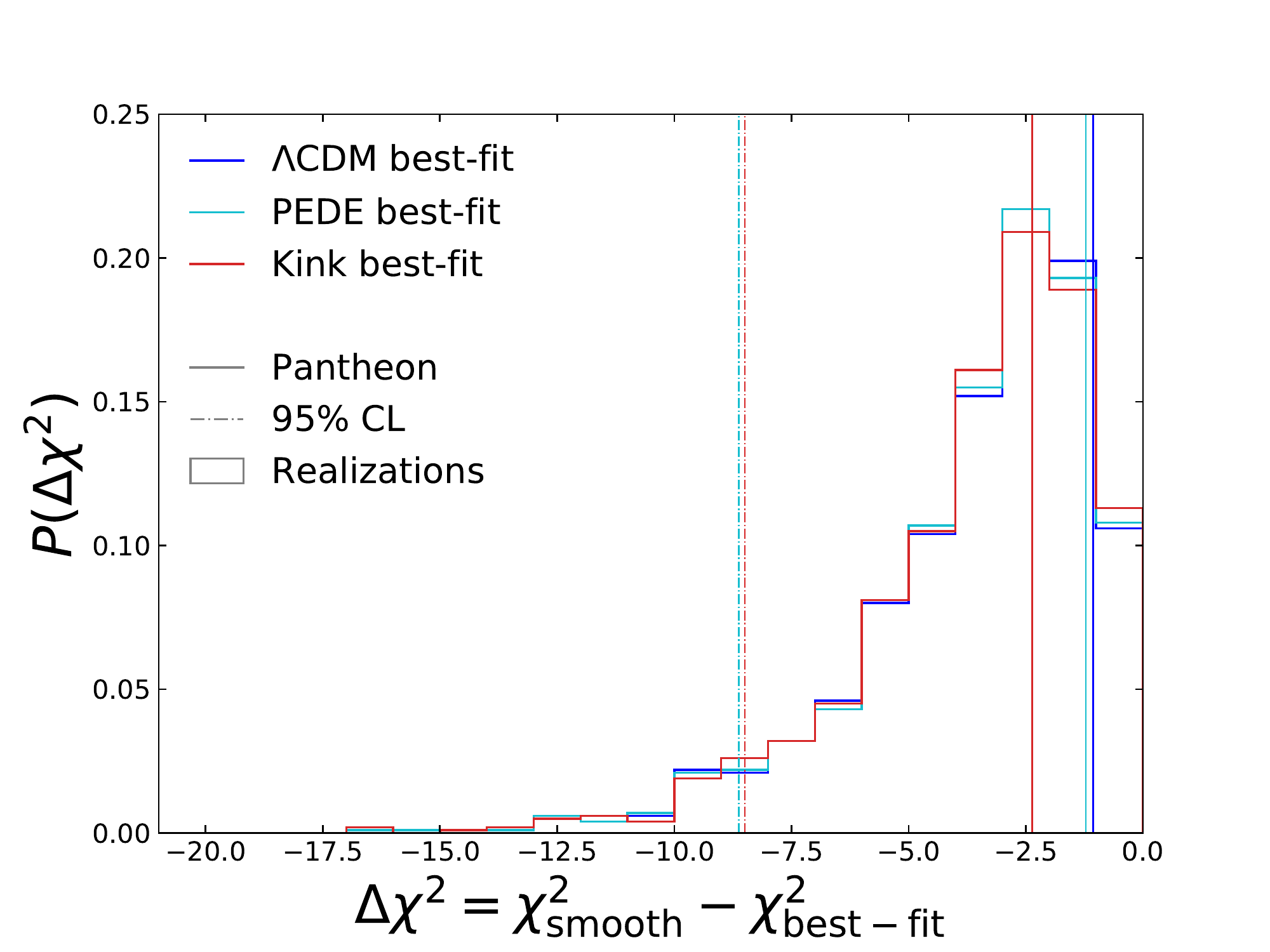}
\caption{Likelihood distributions of $\Delta\chi^2$ for the $\LCDM$, PEDE, and Kink models. The Pantheon-like mock realizations generated from each model were fit with that same model. Thus, the likelihood distributions are the same as each other. The dashed vertical lines correspond to the $\Delta \chi^2$ value such that only 5\% of the realizations have a larger $\Delta \chi^2$ value. The solid vertical lines correspond to the $\Delta \chi^2$ value of the actual Pantheon dataset.}
\label{fig:dchi2_bf_pantheon}
\end{figure}

With our mock datasets in hand, we can then, for each realization and model dataset, find the best-fit parameters of the models for those datasets. The $\chi^2$ of the best-fit parameters, we call $\chi^2_{\rm best-fit}$. Using the distance moduli from the best-fit parameters, we can then start the smoothing procedure  and calculate the $\chi^2$ that results, which we call $\chi^2_{\rm smooth}$. The distribution of the difference between these two $\Delta \chi^2 = \chi^2_{\rm smooth} - \chi^2_{\rm best-fit}$ is our likelihood distribution and we plot these results in Fig.~\ref{fig:dchi2_bf_pantheon}.

With the likelihood distribution, we can then answer how often the smoothing procedure will generate a $\chi^2$ value better than the best-fit model purely by random chance.  Specifically, we find that 95\% of the time, the smoothing procedure will generate a $\Delta \chi^2 \gtrsim -8.6$.
We call this the model's $\Delta \chi^2_{95\%}$.  
The exact values of each model's $\Delta \chi^2_{95\%}$ 
are given in Table \ref{tab:dchi2_bf_pantheon} along with the corresponding actual values for the three considered models fit to the actual Pantheon data. 

With this number in hand, we can perform the smoothing procedure for the actual Pantheon dataset and compare the resulting $\Delta \chi^2$ to this number.  For the actual Pantheon dataset, the smoothing procedure only improves the fit by between $\Delta \chi^2 \sim -1$ and $-2.4$ for any of the considered models.
While the likelihood distribution of the Pantheon data allows the $\Delta \chi^2$ between zero and $-8.6$ at $95\%$, we can see that all of the considered models are consistent with the data. One crucial point to emphasise here is that in this approach, the consistency of a model with the data is tested independent of any alternative model. 
Table~\ref{tab:dchi2_bf_pantheon} shows that the derived $\Delta\chi^2_{95\%}$ is identical for all considered models which shows the reliability of the likelihood distribution.     

\begin{deluxetable}{ccc}
\tablewidth{0pt}
\setlength{\tabcolsep}{5mm}
\tablecaption{\label{tab:dchi2_bf_pantheon}
(Middle) The $\Delta\chi^2$ that corresponds to the 95\% CL for the $\LCDM$, PEDE, and Kink cases.
The three cases yield similar values for the 95\% CLs. (Right) $\Delta\chi^2$ values between the smoothed $\chi^2$ and the three model best-fits to the actual Pantheon data. They are within the 95\% CLs.}
\tablehead{Initial guess & $\Delta\chi^2_{95\%}$ & $\Delta\chi^2_{\rm Pantheon}$ }
\startdata
$\LCDM$ best-fit & -8.63 & -1.06 \\
PEDE best-fit & -8.62 & -1.22 \\
Kink best-fit & -8.50 & -2.36
\enddata
\end{deluxetable}

\newpage
\subsection{Parameter Estimation}

\begin{figure}
\centering
\includegraphics[width=0.45\textwidth]{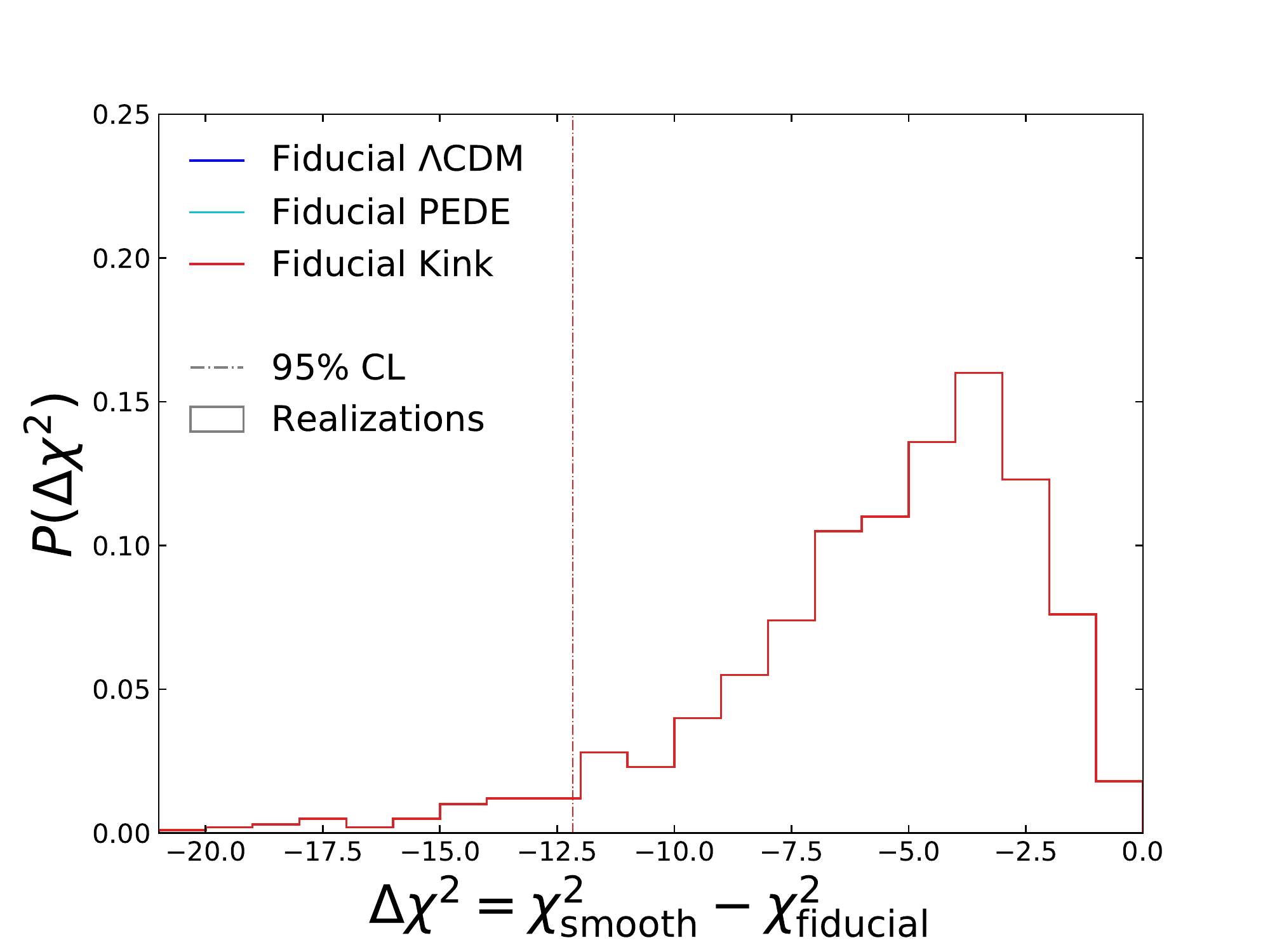}
\caption{Identical likelihood distributions of $\Delta\chi^2$ from Pantheon-like mock realizations, generated from the fiducial $\LCDM$, PEDE, and Kink models. The vertical dashed line show the $\Delta \chi^2$ value that encloses 95\% of the realizations (95\% confidence level).}
\label{fig:dchi2_fid_pantheon}
\end{figure}

\begin{deluxetable}{cc}
\tablewidth{0pt}
\setlength{\tabcolsep}{5mm}
\tablecaption{95\% CLs of $\Delta\chi^2$ from Pantheon-like mock datasets, generated from the fiducial $\LCDM$, PEDE, and Kink models. The three fiducial models give identical values of 95\% CLs.}
\tablehead{Initial guess & $\Delta\chi^2_{95\%}$}
\label{tab:dchi2_fid_pantheon} 
\startdata
Fiducial $\LCDM$ & -12.17 \\
Fiducial PEDE & -12.17 \\
Fiducial Kink & -12.17 
\enddata
\end{deluxetable}

After the initial stages of defining the likelihood distributions and performing model selection, we can do parameter estimation for the models that have shown consistency with the data.
There is another way to define the likelihood distribution that can be useful for the purpose of parameter estimation. 
Due to flexibilities and the existence of free parameters in every cosmological model (e.g. the matter density, curvature, or dark energy equation of state), even if we know the true model of the Universe, the best-fit point would be different from the actual true point in the parameter space of the model. 
In this work, while we assume there is no curvature and fix each model's dark energy evolution, still, the matter density is a free parameter that is fit for. Considering both of these points, we define an alternative likelihood distribution,  $\Delta \chi^2 = \chi^2_{\rm smooth}- \chi^2_{\rm fiducial}$. In this alternative definition, instead of using the likelihood of the best-fit model, we use the likelihood of the true fiducial model and the rest of the procedure is similar to the previous case. 
It is necessary to use the true fiducial point in defining the likelihood distribution since using the best-fit point would result in rejecting the true parameter value at 95\% confidence more than 5\% of the time (1-95\%). 

For the purpose of parameter estimation we use the second likelihood distribution that we derived $\Delta \chi^2 = \chi^2_{\rm smooth}- \chi^2_{\rm fiducial}$ and consider every point in the parameter space of a model as a specific model to be tested individually. 
As an example, a specific model with $\Lambda$ dark energy and $\Omega_{\rm m}=0.3$ would have its own expansion history and likelihood. We can use this point in the parameter space of $\LCDM$ as the initial guess in the smoothing procedure and compare the resulting smoothed $\chi^2$ to that parameter's likelihood. Hence we can perform this approach testing every single individual point in the parameter space of a model and see which ones are consistent with the data at a certain confidence. 

Fig.~\ref{fig:dchi2_fid_pantheon} shows that the derived likelihood distributions for the Pantheon data has a shape independent of the choice of dark energy model that we used to perform the simulations. This is very much expected since the iterative smoothing method deals with residuals and subtracting any true model from its own data realizations would result to the same random residuals independent of the assumed model. The vertical line in this figure show the $\Delta \chi^2 = -12.17$ corresponding to $95 \%$ confidence level. In other words, a true model with $95 \%$ probability would have a $\Delta \chi^2$ better than -12.17 (with respect to the reconstruction from smoothing method). 

We can see that the shape of the likelihood distributions $\Delta \chi^2 = \chi^2_{\rm smooth}- \chi^2_{\rm best-fit}$ are shrunk in comparison to the case of $\Delta \chi^2 = \chi^2_{\rm smooth}- \chi^2_{\rm fiducial}$ and as we explained earlier, this is due to the fact that the best-fit models always have a better likelihood than the true models. 

\begin{figure}
\centering
\includegraphics[width=0.45\textwidth]{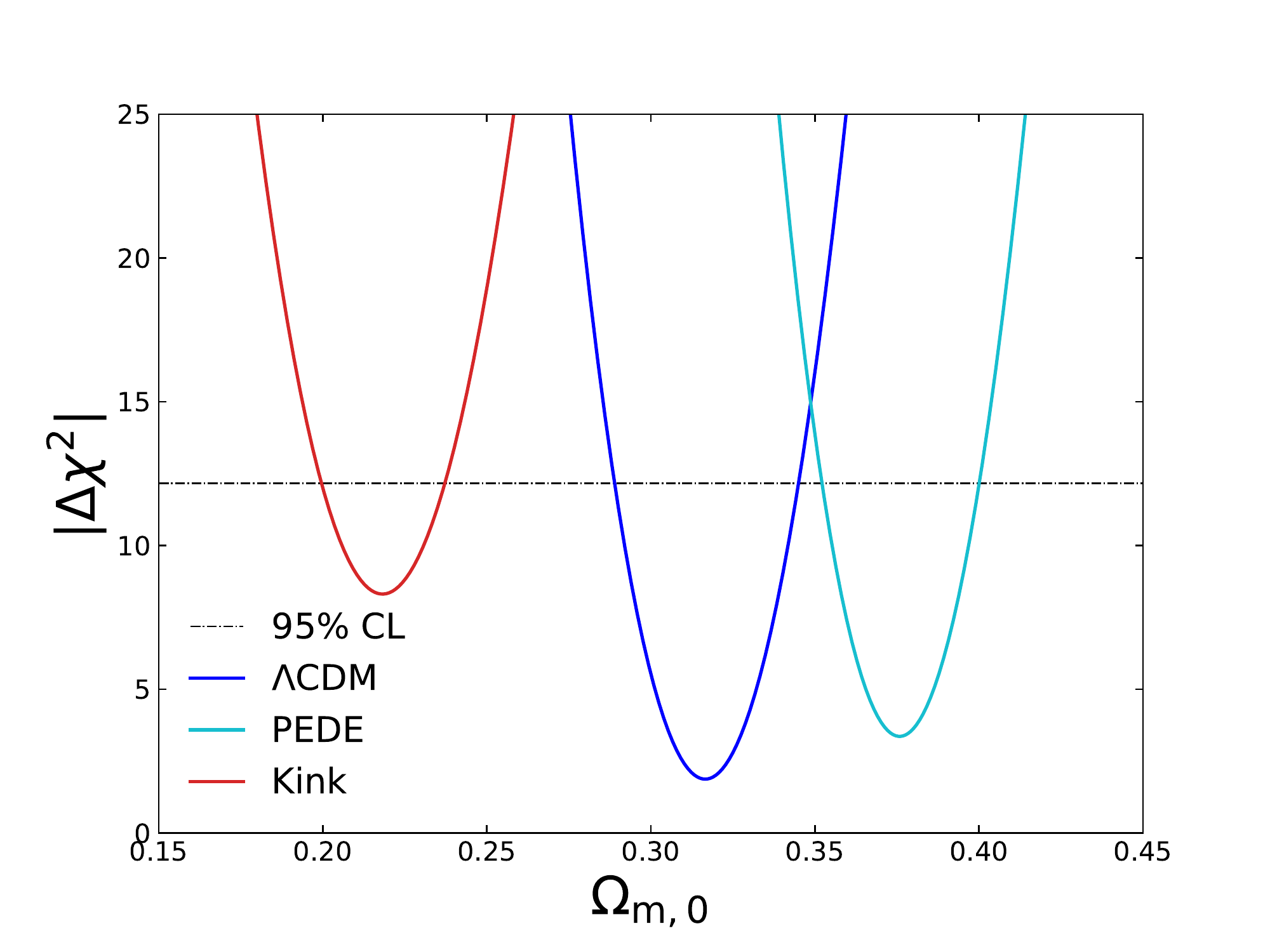}
\caption{\label{fig:dchi2_omegam_pantheon} 
The $\chi^2$ value as a function of matter density for the considered models ($\Lambda$CDM (blue), PEDE (cyan), the Kink model (red)) relative to the smoothed $\chi^2$ value. 
The horizontal dashed line corresponds to $\Delta\chi^2_{95\%}=-12.17$, the value that encloses 95\% of the mock datasets.
}
\end{figure}

Fig.~\ref{fig:dchi2_omegam_pantheon} and Table~\ref{tab:omegam_fid_pantheon} show the $95\%$ CL for the accepted values of matter density for each dark energy model. These values of matter densities for their respected dark energy models, would fit the data with a likelihood that falls within the $95\%$ confidence level derived from our likelihood distribution using many simulations. As one can see, the Pantheon data is consistent with a broader range of matter density for the case of the $\LCDM$ model and interestingly the valid range of matter density for these three models do not overlap at the $95\%$ CL. 

\begin{deluxetable}{cc}
\tablewidth{0pt}
\setlength{\tabcolsep}{5mm}
\tablecaption{95\% confidence interval of $\Omega_{\rm m}$ for the $\LCDM$, PEDE, and Kink models using the $\Delta\chi^2_{95\%}$ value as calculated from the mock realizations.}
\label{tab:omegam_fid_pantheon}
\tablehead{95\% CL & $\Omega_{\rm m}$}
\startdata
$\LCDM$ & $0.289<\Omega_{\rm m}<0.345$ \\
PEDE & $0.352<\Omega_{\rm m}<0.400$ \\
Kink & $0.200<\Omega_{\rm m}<0.237$
\enddata
\end{deluxetable}

\newpage
\section{Type II errors and Forecasting Future Data}\label{sec:wfirst}
\begin{figure}
\centering
\includegraphics[width=0.45\textwidth]{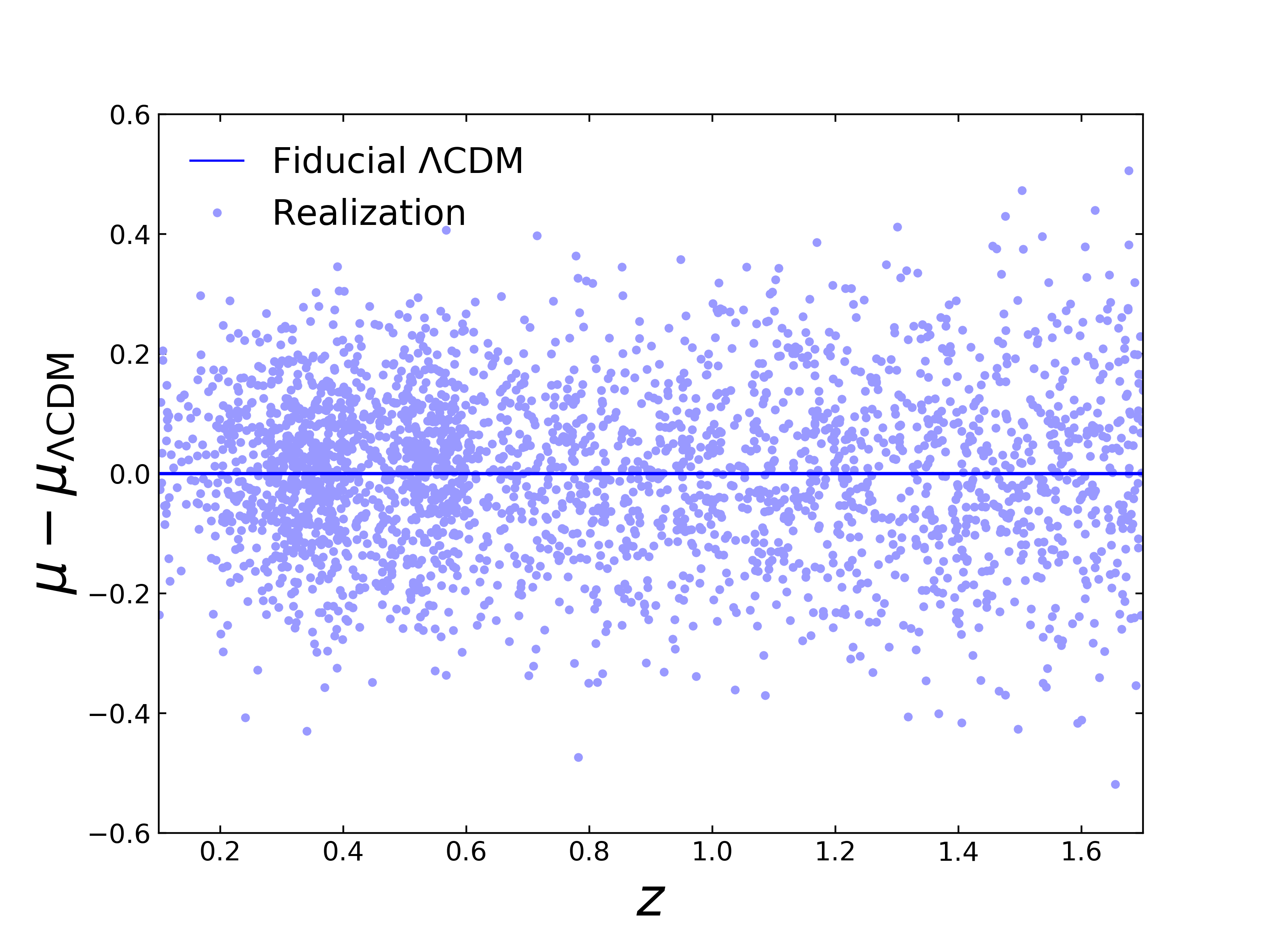}
\caption{\label{fig:dmu_mock_wfirst} One of the 1,000 random mock WFIRST SN Ia compilations based on a fiducial $\LCDM$ model.}
\end{figure}

In this section, we discuss the rate of type II errors for our methodology.  That is, if we make mock datasets from a $\LCDM$ cosmology, how often would we fail to reject the false PEDE or Kink models. 
To answer this question, we forecast the results of our analysis for future WFIRST~\citep{spergel-etal15} data.
We simulate 1000 realizations of the future expected WFIRST data for 
the $\LCDM$ model (with parameter values of $H_0=70 \ \rm{km \ s^{-1} \ Mpc^{-1}}$ and $\Omega_{\rm m}=0.3$) 
as the fiducial model and derive the likelihood distribution  $\Delta \chi^2 = \chi^2_{\rm smooth} - \chi^2_{\rm best-fit}$ as we did in Sec.~2.1 We should emphasize here that this likelihood distribution would be independent of the assumed model in the simulation as we demonstrated earlier in this paper. WFIRST data would provide us with 2725 data points up to redshift of $1.7$. Fig.~\ref{fig:dmu_mock_wfirst} shows one realization of the simulated WFIRST data. We can see that the data can cover a broad redshift range with a high density of the data. 

\begin{figure}
\centering
\includegraphics[width=0.45\textwidth]{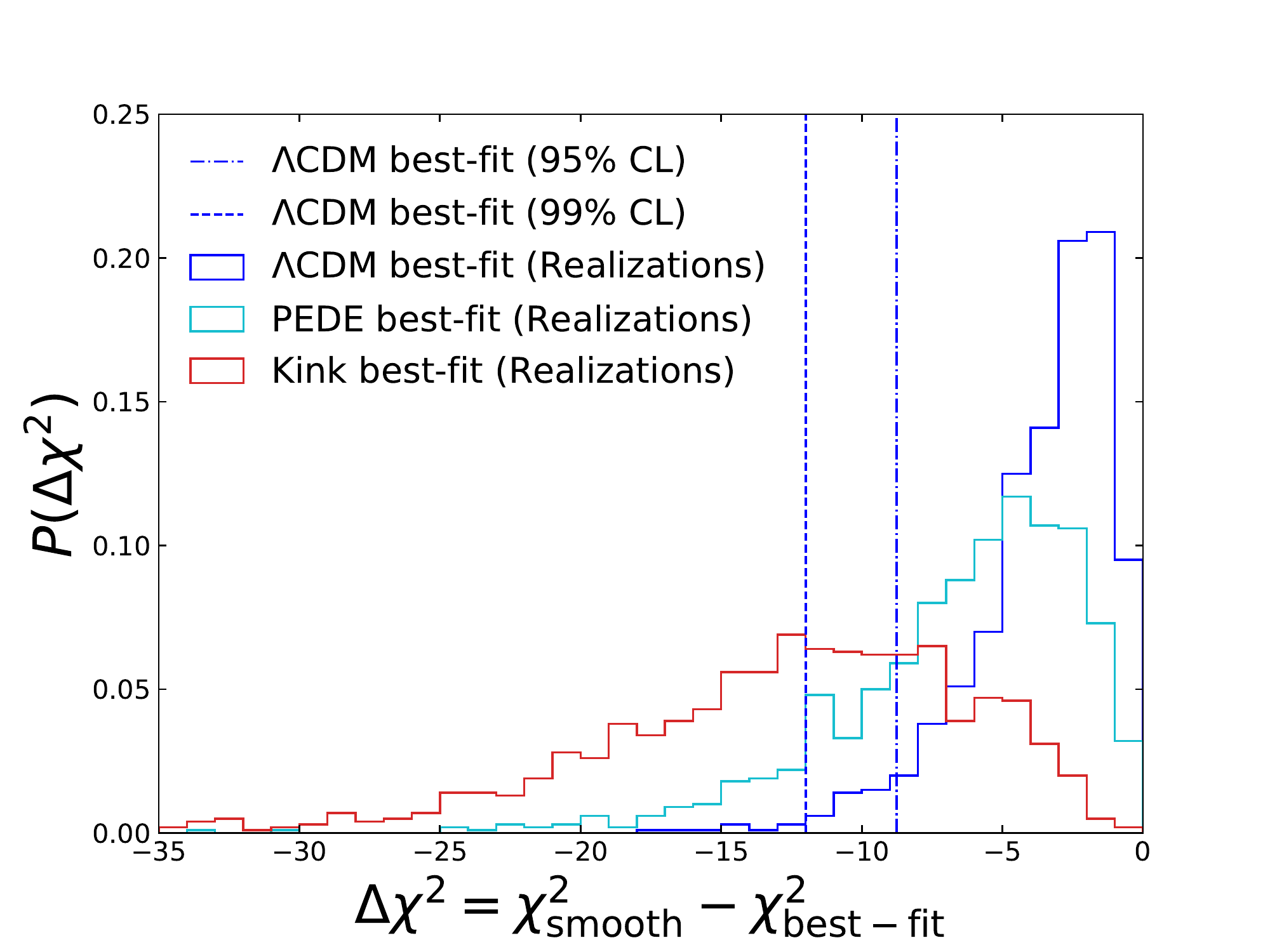}
\caption{Likelihood distributions of $\Delta\chi^2 = \chi^2_{\rm smooth} - \chi^2_{\rm best-fit}$ for each of the considered models, $\LCDM$, PEDE and Kink.  The distribution is over the different realizations of mock WFIRST datasets based on a fiducial $\LCDM$ model. 
\label{fig:dchi2_bf_wfirst}}
\end{figure}

Having the likelihood distribution for the forecasted WFIRST data, and knowing that each realization of the data is based on the $\LCDM$ model, we fit our three considered models, $\LCDM$, PEDE and Kink model, to each realization of the data, and perform the smoothing procedure to calculate the $\Delta \chi^2 = \chi^2_{\rm smooth}- \chi^2_{\rm best-fit}$. Then we look how often the PEDE and Kink likelihood distributions are outside $95\%$ and $99\%$ CLs of the $\LCDM$ likelihood distribution. The results are shown in Fig.~\ref{fig:dchi2_bf_wfirst}. The blue distribution represents the likelihood distribution when the assumed model and the simulated data are both $\LCDM$ model with $\Omega_{\rm m}=0.3$. The two vertical lines represent the $95\%$ and $99\%$ confidence limits. The red and cyan lines represent the likelihood distributions calculated using the best-fit Kink and PEDE models as initial guesses in the smoothing procedure.  We find that, with a type I error rate of 5\% the PEDE model would cause a type II error rate of 75.3\% and the Kink model would cause a type II error rate of 29.9\%. 

We find that, for future WFIRST datasets, in a large number of cases (realizations of the data), the wrong models are now ruled out at $95\%$ and $99\%$ respectively. Table~\ref{tab:nconf_bf95_wfirst} shows in how many cases (realizations of the data), the wrong assumed model (Kink or PEDE) would be outside of the $95\%$ and $99\%$ CL.
We forecast that we can rule out the PEDE model at $95\%$ confidence $24.7\%$ of the time and at $99\%$ confidence, $10.5\%$ of the time. Constraints are tighter for the case of Kink model as we can rule out this model with $95\%$ confidence $70.1\%$ of the time, and with $99\%$ confidence, $49.5\%$ of the time. 
In other words, using this approach and considering future WFIRST data, there is a $70.1\%$ probability that we can rule out this Kink model with more than $95\%$ confidence. We should note here that the tight constraints on cosmological parameters are usually derived by using combination of different data (to break degeneracies) and here we are limiting ourselves to only one type of data for clear demonstration of the approach we are proposing. 

\begin{deluxetable}{ccc}
\tablewidth{0pt}
\setlength{\tabcolsep}{5mm}
\tablecaption{The number of type II successes where the iterative smoothing method was able to accurately rule out the Kink and PEDE models from mock $\LCDM$ data.  The type II error rate is $1-N/1000$ where N are the entries in this table.
\label{tab:nconf_bf95_wfirst} }
\tablehead{ $\LCDM$ data & $N(>95\%\ {\rm CL})$ & $N(>99\%\ {\rm CL})$ }
\startdata
PEDE    & 247 & 105  \\
Kink    & 701 & 495
\enddata
\end{deluxetable}

\section{Summary and Discussion}\label{sec:dis}

We introduce a frequentist test that employs the iterative smoothing method to answer whether a model is a good fit to the data, independent of a comparison with other models.  This works by calculating the likelihood distribution, the distribution of the difference between the $\chi^2$ value produced by the iterative smoothing method and the $\chi^2$ of the best-fit model, for different mock realizations of the data.
We then determine the value of $\Delta \chi^2$ that encloses 95\% and 99\% of the volume of this distribution. 
For our three chosen models with different dark energy properties, we show that the likelihood distributions are the same. 
We conclude that the likelihood distribution is independent of its background model. Thus, we can use this $\Delta \chi^2 = -8.6$ number as a test for real data, even though the true model is unknown.

We find that the $\Delta \chi^2_{95\%}$ for the Pantheon dataset is $-8.6$ while the iterative smoothing method only improves the best-fit of any of the assumed models by less than $-2.4$, thus indicating all of the models are good fits to the data. We also perform parameter estimation for each assumed model which indicate at what values of the matter density, these cosmological models are consistent with the data. Interestingly, the valid ranges of matter density for the three cosmologies we studied do no overlap at 95\% CL. This shows that adding a complimentary data to the analysis one can yield much tighter constrains on the model parameters.

Considering the future data, WFIRST should have enough SN Ia at high redshift to be able to distinguish these models confidently. For example, the Kink model can be ruled out at $>99\%$ confidence in 50\% of our mock realizations based on the $\LCDM$ model. Also, the analyses using WFIRST mock datasets can be done in the same way for forecasting results from other future SN Ia compilations, such as the ones from Dark Energy Spectroscopic Instrument \citep{desi-etal16a,desi-etal16b} and Large Synoptic Survey Telescope \citep{ivezic-etal19}. These surveys may help us to detect any possible deviation from the standard $\LCDM$ cosmological model.

In the next companion paper we will compare the power of our approach in model selection and parameter estimation with the conventional approach based on Bayesian evidence.

\acknowledgments

This work was supported by the high performance computing clusters Seondeok at the Korea Astronomy and Space Science Institute. A.~S. would like to acknowledge the support of the Korea Institute for Advanced Study (KIAS) grant funded by the government of Korea. B.~L. would like to acknowledge the support of the National Research Foundation of Korea (NRF-2019R1I1A1A01063740).

\bibliography{KSKL20-2}

\end{document}